\documentclass[aps,prd,twocolumn,showpacs,showkeys,preprintnumbers,amsmath,amssymb,floatfix,longbibliography,nofootinbib]{revtex4-2}
\pdfoutput=1
 
\usepackage{amsmath,latexsym}
\usepackage{xcolor}
\usepackage[colorlinks=true, urlcolor=blue, linkcolor=blue, citecolor=blue]{hyperref}
\usepackage{etoolbox}
\usepackage{breqn}

\usepackage{graphicx}
\usepackage{dcolumn}
\usepackage{bm}

\def\beq{\begin{equation}}
\def\eeq{\end{equation}}
\def\beqn{\begin{eqnarray}}
\def\eeqn{\end{eqnarray}}
\def\eeqno#1{\label{#1}\end{equation}}

\newcommand{\be}{\begin{equation}}
\newcommand{\ee}{\end{equation}}
\newcommand{\bea}{\begin{eqnarray}}
\newcommand{\eea}{\end{eqnarray}}

\makeglossary
\makeindex

\begin{document}

\preprint{JGM/016-The Isomorphism of 3-Qubit Hadamards and E8}
\title[The Isomorphism of 3-Qubit Hadamards and $E_8$]{The Isomorphism of 3-Qubit Hadamards and $E_8$}
%\dedicatory{Dedicated to the pursuit of Truth and beauty in Nature}
\url{https://www.TheoryOfEverything.org}

\author{J Gregory Moxness}
\homepage{https://www.TheoryOfEverything.org/theToE}
\email[mailto:jgmoxness@TheoryOfEverything.org]{}
\affiliation{TheoryOfEverything.org}

%\date{\today}
\date{November 20,2023}

\begin{abstract}
This paper presents several notable properties of the matrix $\mathbb{U}$ shown to be related to the isomorphism between $H_4$ and $E_8$. The most significant of these properties is that $\mathbb{U}$.$\mathbb{U}$ is to rank 8 matrices what the golden ratio is to numbers. That is to say, the difference between it and its inverse is the identity element, albeit with a twist. Specifically, $\mathbb{U}$.$\mathbb{U}$-$ (\mathbb{U}$.$\mathbb{U})^{-1}$ is the reverse identity matrix or standard involutory permutation matrix of rank 8. It has the same palindromic characteristic polynomial coefficients as the normalized 3-qubit Hadamard matrix with 8-bit binary basis states, which is known to be isomorphic to E8 through its (8,4) Hamming code. 
\end{abstract}


\maketitle

\section{Introduction}

The Split Real Even (SRE) form of the $E_8$ Lie group with a unimodular lattice in $\mathbb{R}^{8}$ has 240 vertices and 6,720 edges of 8-dimensional (8D) length $\sqrt{2}$. $E_8$ is the largest of the exceptional simple Lie algebras, groups, lattices, and polytopes related to octonions ($\mathbb{O}$), (8,4) Hamming codes, and 3-qubit (8 basis state) Hadamard matrix gates. An important and related higher dimensional structure is the $\mathbb{R}^{24}$ ($\mathbb{C}^{12}$) Leech lattice ($\Lambda_{24}\supset$$E_8$$\oplus$$E_8$$\oplus$$E_8$), with its binary (ternary) Golay code construction. 

It has been shown\cite{Moxness2023-015} that the matrix $\mathbb{U}$ in (\ref{eqn:H4foldUnitary}) along with its inverse (\ref{eqn:InvH4foldUnitary}) is related to the isomorphism between $H_4$ and $E_8$. 

\begin{equation}
\label{eqn:H4foldUnitary}
\text{$ \mathbb{ U } $}\text{ = } \left(
\begin{array}{cccccccc}
1- \varphi& 0 & 0 & 0 & 0 & 0 & 0 & -\varphi ^{2} \\
0 & -1 & \varphi & 0 & 0 & \varphi & 1 & 0 \\
0 & \varphi & 0 & 1 & -1 & 0 & \varphi & 0 \\
0 & 0 & -1 & \varphi & \varphi & 1 & 0 & 0 \\
0 & 0 & 1 & \varphi & \varphi & -1 & 0 & 0 \\
0 & \varphi & 0 & 1 & -1 & 0 & \varphi & 0 \\
0 & 1 & \varphi & 0 & 0 & \varphi & -1 & 0 \\
 -\varphi ^{2} & 0 & 0 & 0 & 0 & 0 & 0 &1- \varphi \\
\end{array}
\right)/(2 \sqrt{\varphi})
\end{equation}

\begin{equation}
\label{eqn:InvH4foldUnitary}
\text{$\mathbb{U}^{-1}$}\text{=}\left(
\begin{array}{cccccccc}
\varphi-1& 0 & 0 & 0 & 0 & 0 & 0 & -\varphi ^{2} \\
0 & -\varphi & 1 & 0 & 0 & 1 & \varphi & 0 \\
0 & 1 & 0 & \varphi & -\varphi & 0 & 1 & 0 \\
0 & 0 & -\varphi &1 & 1 & \varphi & 0 & 0 \\
0 & 0 & \varphi & 1 &1& -\varphi & 0 & 0 \\
0 & 1 & 0 & \varphi & -\varphi & 0 & 1 & 0 \\
0 & \varphi & 1 & 0 & 0 & 1 & -\varphi & 0 \\
 -\varphi ^{2} & 0 & 0 & 0 & 0 & 0 & 0 &\varphi-1 \\
\end{array}
\right)/(2 \sqrt{\varphi})
\end{equation}

\begin{figure}[!h]
\center
\includegraphics[width=250pt]{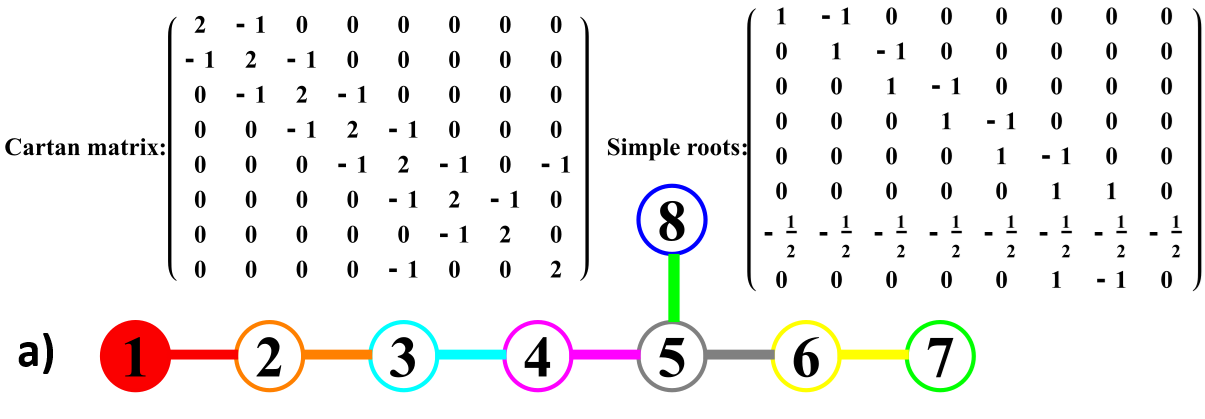}
\caption{\label{fig:421-dynkin} a) $E_8$ Dynkin diagram with it Cartan matrix and simple roots matrix}
\end{figure}

The Coxeter-Dynkin diagram for $E_8$ is shown in Fig. \ref{fig:421-dynkin} along with its Cartan matrix ($\mathtt{cmE8}$) and simple roots matrix ($\mathtt{srE8}$). It has been shown\cite{Moxness2014-006} that $\mathtt{cmE8}$=$\mathtt{srE8}$.$\mathtt{srE8}^{T}$, such that we can think of the simple roots as $\sqrt{\mathtt{cmE8}}$. It was also shown that the SRE $E_8$ vertex coordinates can be derived from the dot product of $\pm\mathtt{ E8roots}$.$\mathtt{srE8}$. Applying these relationships to $\mathbb{U}$ gives interesting results as described in Section \ref{sec:Properties of U}.

\section{Properties of $\mathbb{U}$}
\label{sec:Properties of U}

Similar to the relationships between the Cartan matrix, $\pm$roots, weights, heights of $E_8$, we can construct a Cartan matrix $\mathtt{cm\mathbb{U}}$=$\mathbb{U}$.$\mathbb{U}$ shown in (\ref{eqn:cmU}), with $\mathbb{U}$ playing the role of the simple roots matrix.

Just as the golden ratio $\varphi=\frac{1}{2}\left(1+\sqrt{5}\right)\approx1.618$ generates the integer identity $\varphi$-1/$\varphi$=1, we now have $\mathtt{cm\mathbb{U}}$-$\mathtt{cm\mathbb{U}}^{-1}$ generating the exchange matrix or standard involutory permutation matrix of rank 8 shown in (\ref{eqn:reverseIM8}). This has the same palindromic characteristic polynomial coefficients (\ref{eqn:HadamardCoefficients}) as the normalized 3-qubit Hadamard matrix with 8-bit binary basis states shown in (\ref{eqn:normHad}), which has been shown by Elkies\cite{572561} to be isomorphic to $E_8$ through its (8,4) Hamming code. 

\begin{equation}
\label{eqn:cmU}
\mathtt{cm\mathbb{U}}\text{=}\left(
\begin{array}{cccccccc}
 \frac{\sqrt{5}}{2} & 0 & 0 & 0 & 0 & 0 & 0 & \frac{1}{2} \\
 0 & \frac{\sqrt{5}}{2} & 0 & 0 & 0 & 0 & \frac{1}{2} & 0 \\
 0 & 0 & \frac{\sqrt{5}}{2} & 0 & 0 & \frac{1}{2} & 0 & 0 \\
 0 & 0 & 0 & \frac{\sqrt{5}}{2} & \frac{1}{2} & 0 & 0 & 0 \\
 0 & 0 & 0 & \frac{1}{2} & \frac{\sqrt{5}}{2} & 0 & 0 & 0 \\
 0 & 0 & \frac{1}{2} & 0 & 0 & \frac{\sqrt{5}}{2} & 0 & 0 \\
 0 & \frac{1}{2} & 0 & 0 & 0 & 0 & \frac{\sqrt{5}}{2} & 0 \\
 \frac{1}{2} & 0 & 0 & 0 & 0 & 0 & 0 & \frac{\sqrt{5}}{2} \\
\end{array}
\right)
\end{equation}

\begin{equation}
\label{eqn:reverseIM8}
\mathtt{cm\mathbb{U}}-\mathtt{cm\mathbb{U}}^{-1}\text{=}\left(
\begin{array}{cccccccc}
 0 & 0 & 0 & 0 & 0 & 0 & 0 & 1 \\
 0 & 0 & 0 & 0 & 0 & 0 & 1 & 0 \\
 0 & 0 & 0 & 0 & 0 & 1 & 0 & 0 \\
 0 & 0 & 0 & 0 & 1 & 0 & 0 & 0 \\
 0 & 0 & 0 & 1 & 0 & 0 & 0 & 0 \\
 0 & 0 & 1 & 0 & 0 & 0 & 0 & 0 \\
 0 & 1 & 0 & 0 & 0 & 0 & 0 & 0 \\
 1 & 0 & 0 & 0 & 0 & 0 & 0 & 0 \\
\end{array}
\right)
\end{equation}

\begin{equation}
\label{eqn:normHad}
H\text{=}\left(
\begin{array}{cccccccc}
 1 & 1 & 1 & 1 & 1 & 1 & 1 & 1 \\
 1 & -1 & 1 & -1 & 1 & -1 & 1 & -1 \\
 1 & 1 & -1 & -1 & 1 & 1 & -1 & -1 \\
 1 & -1 & -1 & 1 & 1 & -1 & -1 & 1 \\
 1 & 1 & 1 & 1 & -1 & -1 & -1 & -1 \\
 1 & -1 & 1 & -1 & -1 & 1 & -1 & 1 \\
 1 & 1 & -1 & -1 & -1 & -1 & 1 & 1 \\
 1 & -1 & -1 & 1 & -1 & 1 & 1 & -1 \\
\end{array}
\right)/\sqrt{8}
\end{equation}

\begin{figure}[!ht]
\center
\includegraphics[width=110pt]{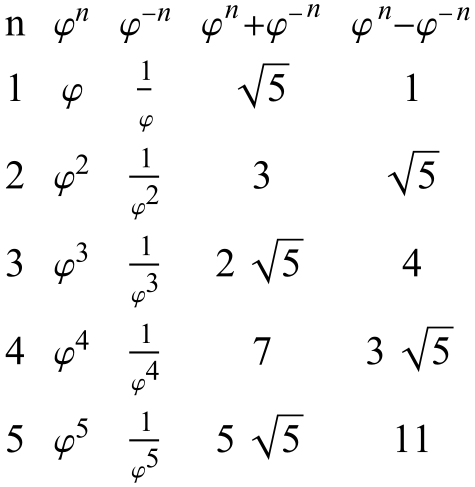}
\caption{\label{fig:phi-powers}Sum and difference in powers of $\varphi$}
\end{figure}

Just as $\frac{\varphi+1/\varphi}{2\varphi-1}$=1, $\frac{\mathtt{cm\mathbb{U}}+\mathtt{cm\mathbb{U}}^{-1}}{2\varphi-1}$=8x8 Identity Matrix. Of course, we can reverse the rows in $\mathtt{cm\mathbb{U}}$, which then swaps the sum and difference operation results of Identity vs. Involutory permutation matrices (respectively). Also as the exponentiation of sum (difference) $\varphi^n\pm 1/\varphi^n$ results in integer factors on even (odd) n and integer radicand factors on odd (even) n as shown in Fig. \ref{fig:phi-powers}, by using matrix power operations on $\mathtt{cm\mathbb{U}^n}\pm \mathtt{cm\mathbb{U}}^{-n}$ produces the Identity (Involutory) matrices with those same scaling factors. This application of matrix powers to $\mathtt{\mathbb{U}}$ instead of $\mathtt{cm\mathbb{U}}$ puts all even n as the alternating integer (integer radicand) matrices, with odd n shown in (\ref{eqn:U-odd-powers-plus}) and (\ref{eqn:U-odd-powers-minus}). Please note that like the 3-qubit Hadamard, inside the square brackets these matrices are traceless and unitary with the characteristic polynomial of (\ref{eqn:HadamardCoefficients}). 

\begin{multline}
\label{eqn:U-odd-powers-plus}
\mathbb{U}^{odd(n)}+\mathbb{U}^{-odd(n)}\text{=} \\
-\left[\frac{1}{2} \left(
\begin{array}{cccccccc}
 0 & 0 & 0 & 0 & 0 & 0 & 0 & 2 \\
 0 & 1 & -1 & 0 & 0 & -1 & -1 & 0 \\
 0 & -1 & 0 & 1 & -1 & 0 & -1 & 0 \\
 0 & 0 & 1 & -1 & -1 & -1 & 0 & 0 \\
 0 & 0 & -1 & -1 & -1 & 1 & 0 & 0 \\
 0 & -1 & 0 & -1 & 1 & 0 & -1 & 0 \\
 0 & -1 & -1 & 0 & 0 & -1 & 1 & 0 \\
 2 & 0 & 0 & 0 & 0 & 0 & 0 & 0 \\
\end{array}
\right)\right]\frac{\varphi ^n+1}{ \varphi ^{n/2}}
\end{multline}

\begin{multline}
\label{eqn:U-odd-powers-minus}
\mathbb{U}^{odd(n)}-\mathbb{U}^{-odd(n)}\text{=} \\
-\left[\frac{1}{2} \left(
\begin{array}{cccccccc}
 2 & 0 & 0 & 0 & 0 & 0 & 0 & 0 \\
 0 & -1 & -1 & 0 & 0 & -1 & 1 & 0 \\
 0 & -1 & 0 & -1 & 1 & 0 & -1 & 0 \\
 0 & 0 & -1 & -1 & -1 & 1 & 0 & 0 \\
 0 & 0 & 1 & -1 & -1 & -1 & 0 & 0 \\
 0 & -1 & 0 & 1 & -1 & 0 & -1 & 0 \\
 0 & 1 & -1 & 0 & 0 & -1 & -1 & 0 \\
 0 & 0 & 0 & 0 & 0 & 0 & 0 & 2 \\
\end{array}
\right)\right]\frac{\varphi ^n-1}{ \varphi ^{n/2}}
\end{multline}

From \cite{Moxness2023-015} we know that $\mathbb{U}$ produces the folding of $E_8$ to $H_4$ with $\mathbb{U}^{-1}$ involved in the unfolding back to $E_8$. We also know its palindromic characteristic polynomial coefficients are those shown in (\ref{eqn:UCoefficients}) with the same form as (\ref{eqn:HadamardCoefficients}). This gives us a better understanding of why $E_8$ is isomorphic to both the Hadamard matrix and $H_4$. Given that the sum, difference, product, and division of $\mathbb{U}$ and cm$\mathbb{U}$ generate both the left and right matrix identities of rank 8 suggests a possible connection to Bott periodicity.

\begin{equation}
\label{eqn:HadamardCoefficients}
H_{cp}\text{=}x^8-4 x^6+6 x^4-4 x^2+1
\end{equation}

\begin{equation}
\label{eqn:UCoefficients}
U_{cp}\text{=}x^8-2\sqrt{5} x^6+7 x^4-2\sqrt{5} x^2+1
\end{equation}

Exploring further, if we take seriously the idea of $\mathtt{cm\mathbb{U}}$ as a Cartan matrix, it can be visualized with its positive roots, weights, heights, and Hasse diagrams as shown in Appendix \ref{app:A} Figs. \ref{fig:cmUroots}-\ref{fig:cmUHasse}. After deleting duplicates generated in the SuperLie\cite{Grozman_2004} analysis of $\mathtt{cm\mathbb{U}}$, the cumulative index count up to height 8 is same as that of $E_8$ being 120. 

\begin{figure}[!ht]
\ \\

\center
\includegraphics[width=250pt]{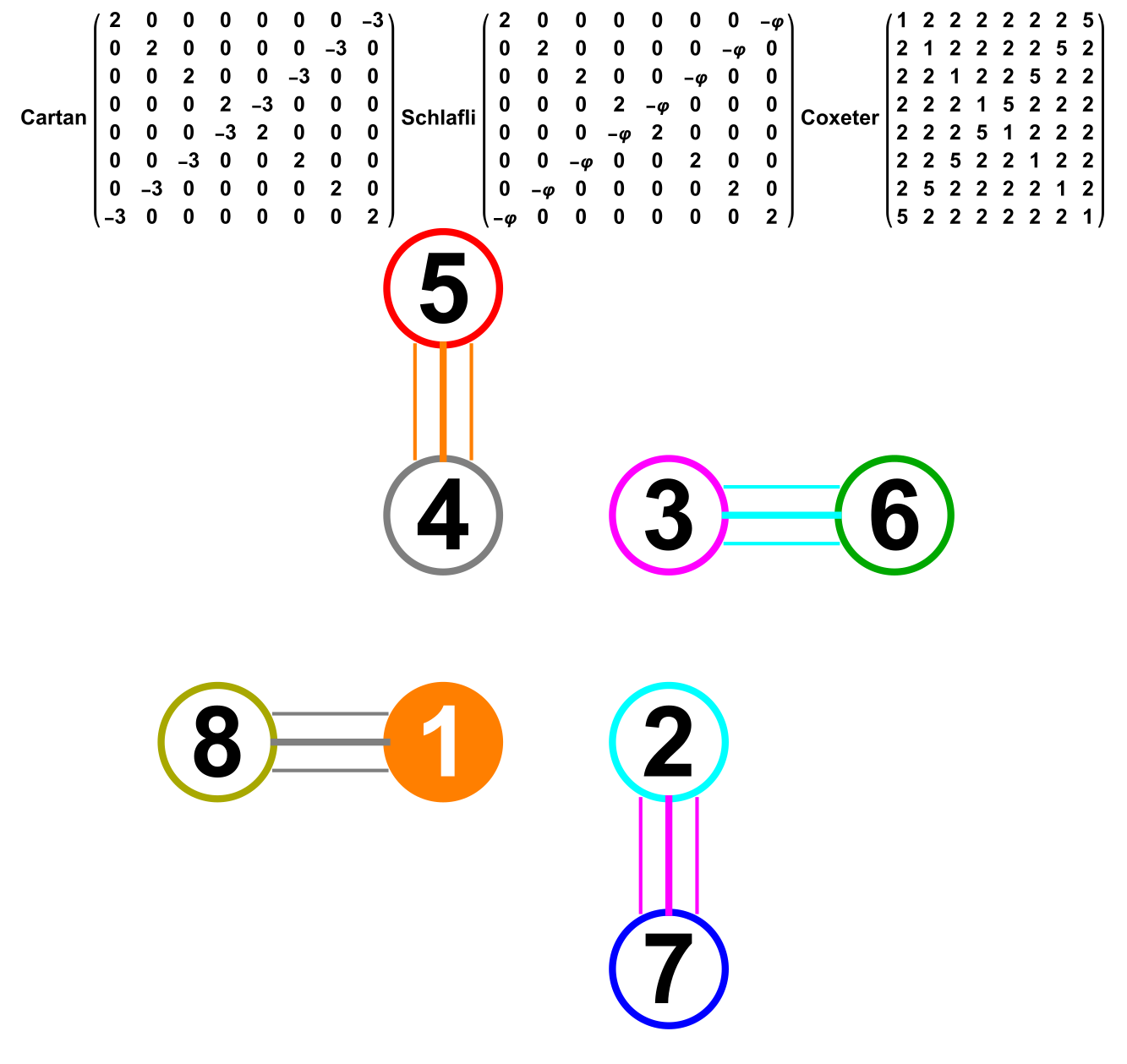}
\caption{\label{fig:U-palindromic-dynkin}$\mathtt{cm\mathbb{U}}$ Coxeter-Dynkin diagram, with Cartan, Schlafli, and Coxeter matrices}
\end{figure}

A corresponding Coxeter-Dynkin diagram, with Cartan, Schlafli, and Coxeter matrices is shown in Fig. \ref{fig:U-palindromic-dynkin}, noting the Schlafli matrix adding fractional 1/2 scaling on the Identity and -3/2 scaling on the Involutory matrices reproduces $-\mathbb{U}^{-1}$. 

If we do the same analysis using the involutory permutation matrix of rank 8 (\ref{eqn:reverseIM8}) as a Cartan matrix, it shows the only difference is the weights are now integers as opposed to factors of $\varphi$. This is shown in Appendix \ref{app:A} Figs. \ref{fig:revIM8roots}-\ref{fig:revIMHasse}.

Generating as prescribed above the $\mathtt{cm\mathbb{U}}$-based vertex coordinates and projecting to 3D using the methods shown in \cite{Moxness2023-015} gives somewhat different results than with the folded or unfolded $E_8$. Instead of finding each of 56 possible subsets of 3 dimensions having the same tally of hull groupings with the same hull geometries, $\mathbb{U}$ groupings rotate into much smaller groups as shown in Fig. \ref{fig:UvertsIn3D}. A more complete hull breakdown using dimensions \{2,3,4\} is shown in Appendix \ref{app:B} Fig. \ref{fig:UvertsInPerms3D-4}, noting the predominance of regular octahedral and irregular icosahedral hulls.

\begin{figure}[!ht]
\center
\includegraphics[width=250pt]{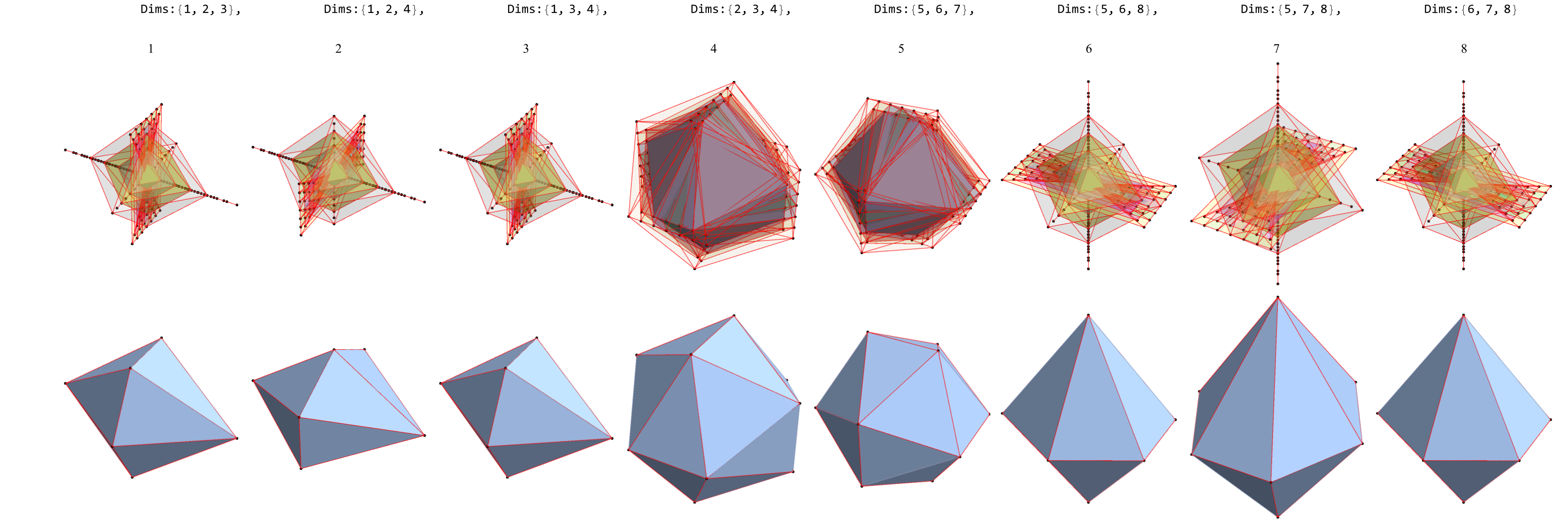}
\caption{\label{fig:UvertsIn3D}Eight orthogonal projections to 3D of $\mathtt{cm\mathbb{U}}$-based vertices from the $\pm$roots of $\mathtt{cm\mathbb{U}}$}
\end{figure}

\section{Conclusion}

This paper presents properties of the matrix $\mathbb{U}$ shown to be related to the isomorphism between $H_4$ and $E_8$. Significantly, $\mathbb{U}$.$\mathbb{U}$ is to rank 8 matrices what the golden ratio is to numbers, such that $\mathbb{U}$.$\mathbb{U}$-$ (\mathbb{U}$.$\mathbb{U})^{-1}$ is the reverse identity matrix or standard involutory permutation matrix of rank 8, with a possible connection to Bott periodicity. It has the same palindromic characteristic polynomial coefficients as the normalized 3-qubit Hadamard matrix with 8-bit binary basis states known to be isomorphic to E8 through its (8,4) Hamming code. In addition to providing insight into the isomorphisms of $E_8$, taking advantage of this property may open the door to as yet unexplored $E_8$-based Grand Unified Theories or GUTs. It is anticipated that these visualizations and connections will be useful in discovering new insights into unifying the mathematical symmetries as they relate to unification in theoretical physics. 

\begin{acknowledgments}

I would like to thank my wife for her love and patience and those in academia who have taken the time to review this work.

\end{acknowledgments}

%\nocite{*}
\bibliography{The_Isomorphism_of_3-Qubit_Hadamards_and_E8}

\appendix

\begin{figure}[!ht]
\center
\includegraphics[width=490pt]{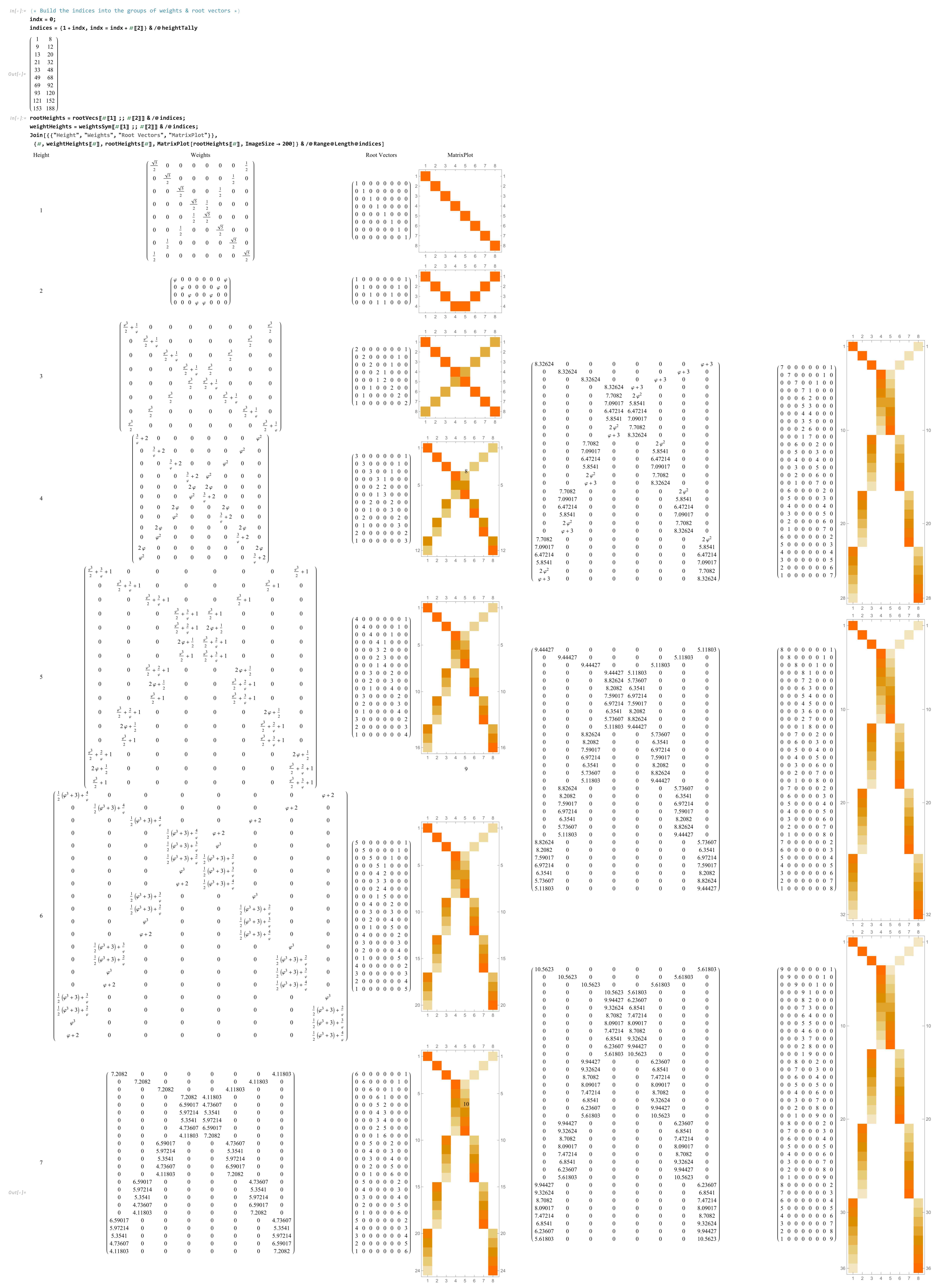}
\caption{\label{fig:cmUroots}Analysis of $\mathtt{cm\mathbb{U}}$ showing the cumulative height group indices, positive roots, weights, and heights\\
Note: At height 8 the cumulative index count is 120, giving 240 positive and negative roots as in $E_8$}
\end{figure}

\begin{figure}[!ht]
\center
\includegraphics[width=525pt]{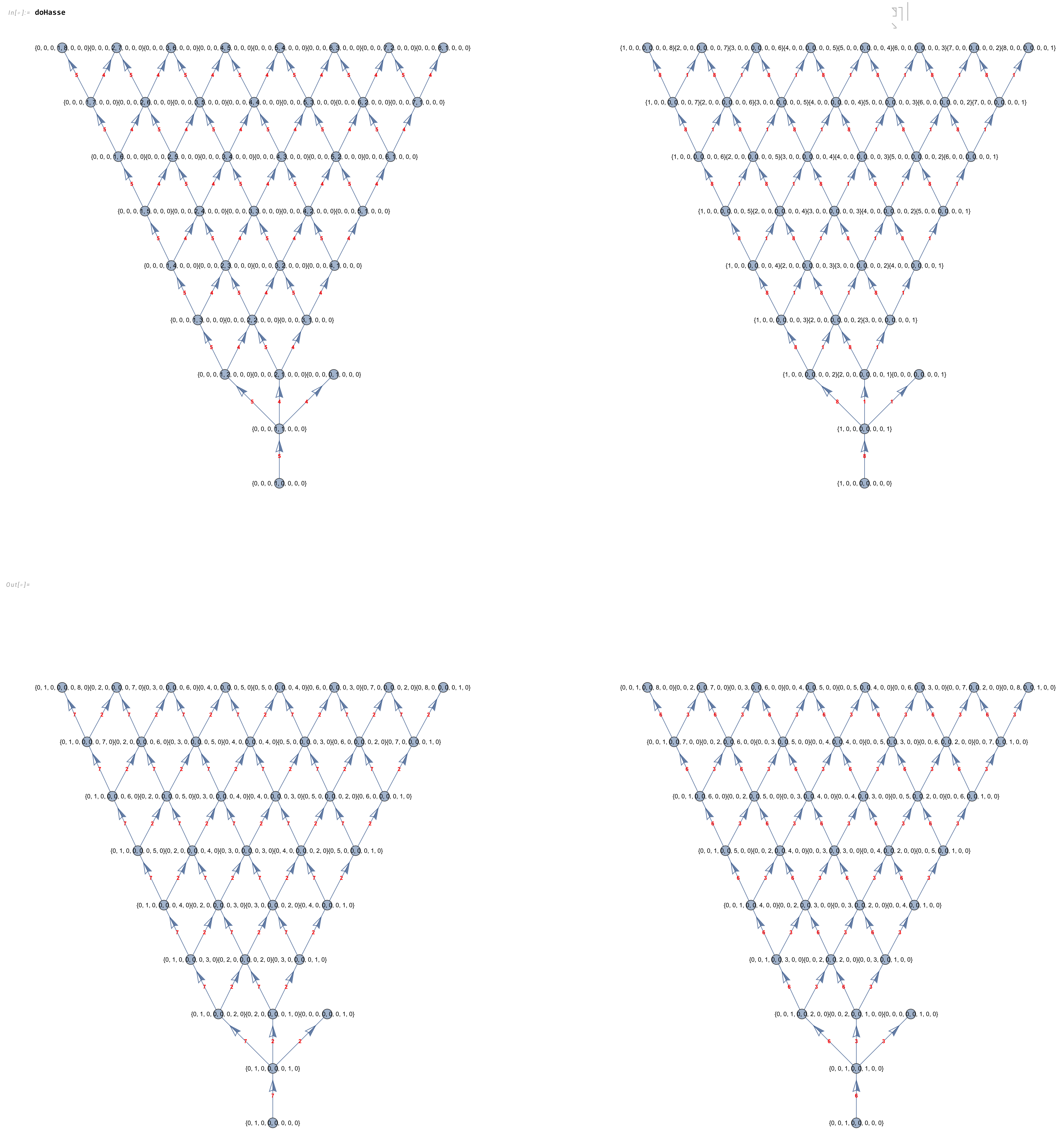}
\caption{\label{fig:cmUHasse}Analysis of $\mathtt{cm\mathbb{U}}$ showing its Hasse visualizations up to height 10}
\end{figure}

\begin{figure}[!ht]
\center
\includegraphics[width=300pt]{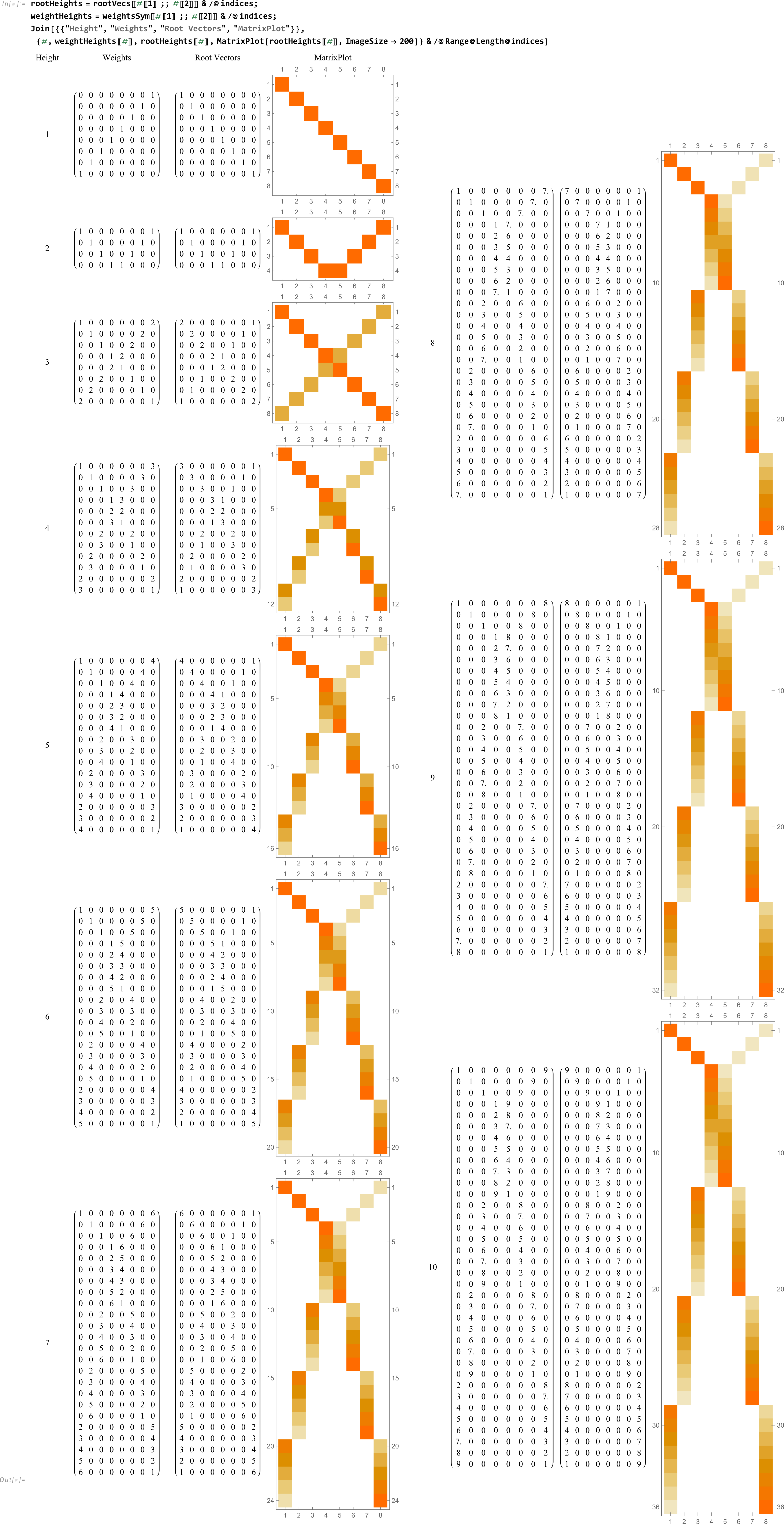}
\caption{\label{fig:revIM8roots}Analysis of $\mathtt{cm\mathbb{U}}$-$\mathtt{cm\mathbb{U}}^{-1}$ showing all integer positive roots, weights, heights}
\end{figure}

\begin{figure}[!ht]
\center
\includegraphics[width=525pt]{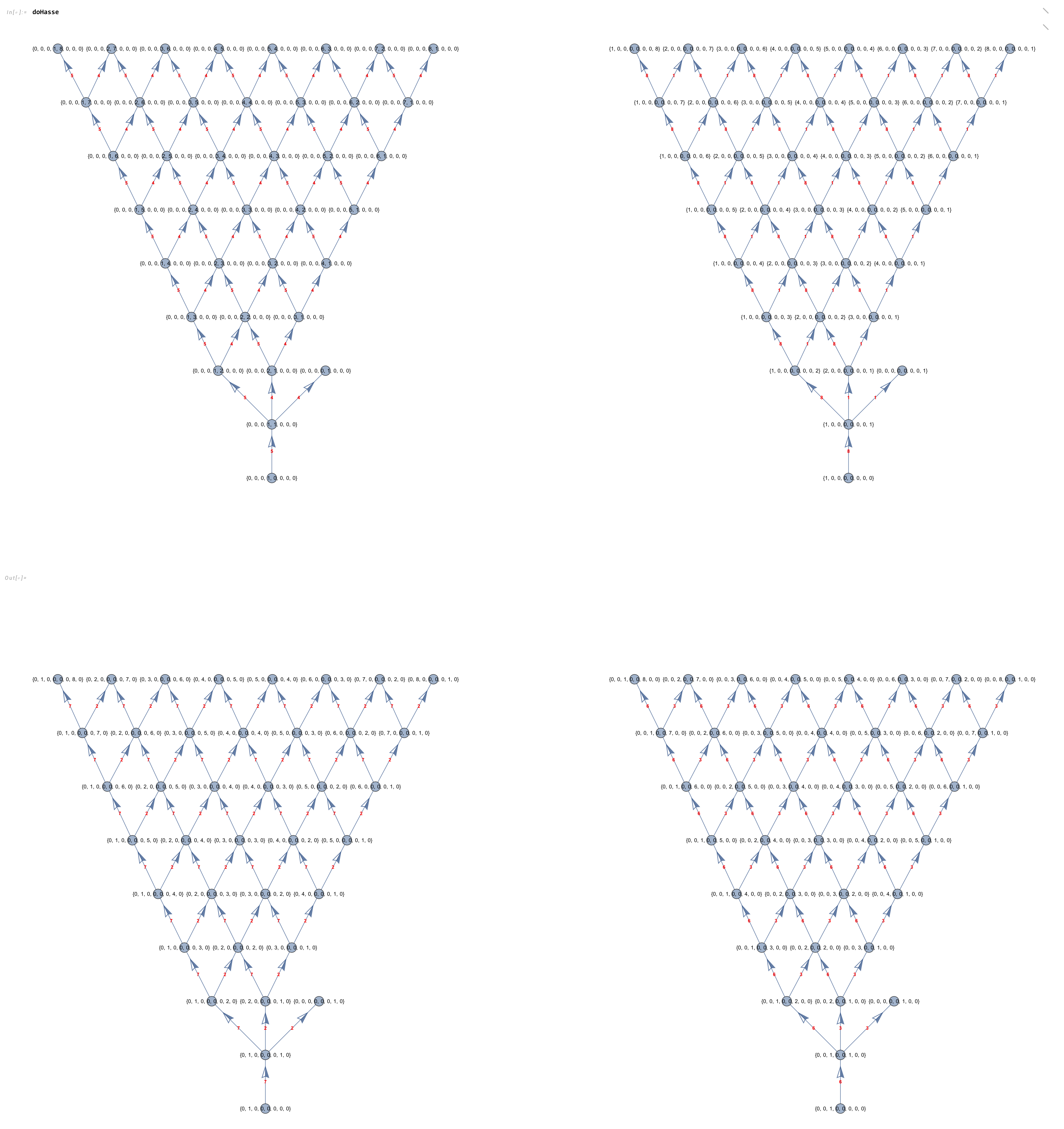}
\caption{\label{fig:revIMHasse}Analysis of $\mathtt{cm\mathbb{U}}$-$\mathtt{cm\mathbb{U}}^{-1}$ showing its Hasse visualizations up to height 10, which are identical to those in Fig. \ref{fig:cmUHasse}}
\end{figure}

\section{\label{app:A}\textit{SuperLie package analysis of $\mathtt{cm\mathbb{U}}$ and the rank 8 involution permutation matrix showing the positive roots, weights, heights, and Hasse visualizations up to height 10}\\
Figs. \ref{fig:cmUroots}-\ref{fig:revIMHasse}
\ \\}

\begin{figure}[!ht]
\center
\includegraphics[width=350pt]{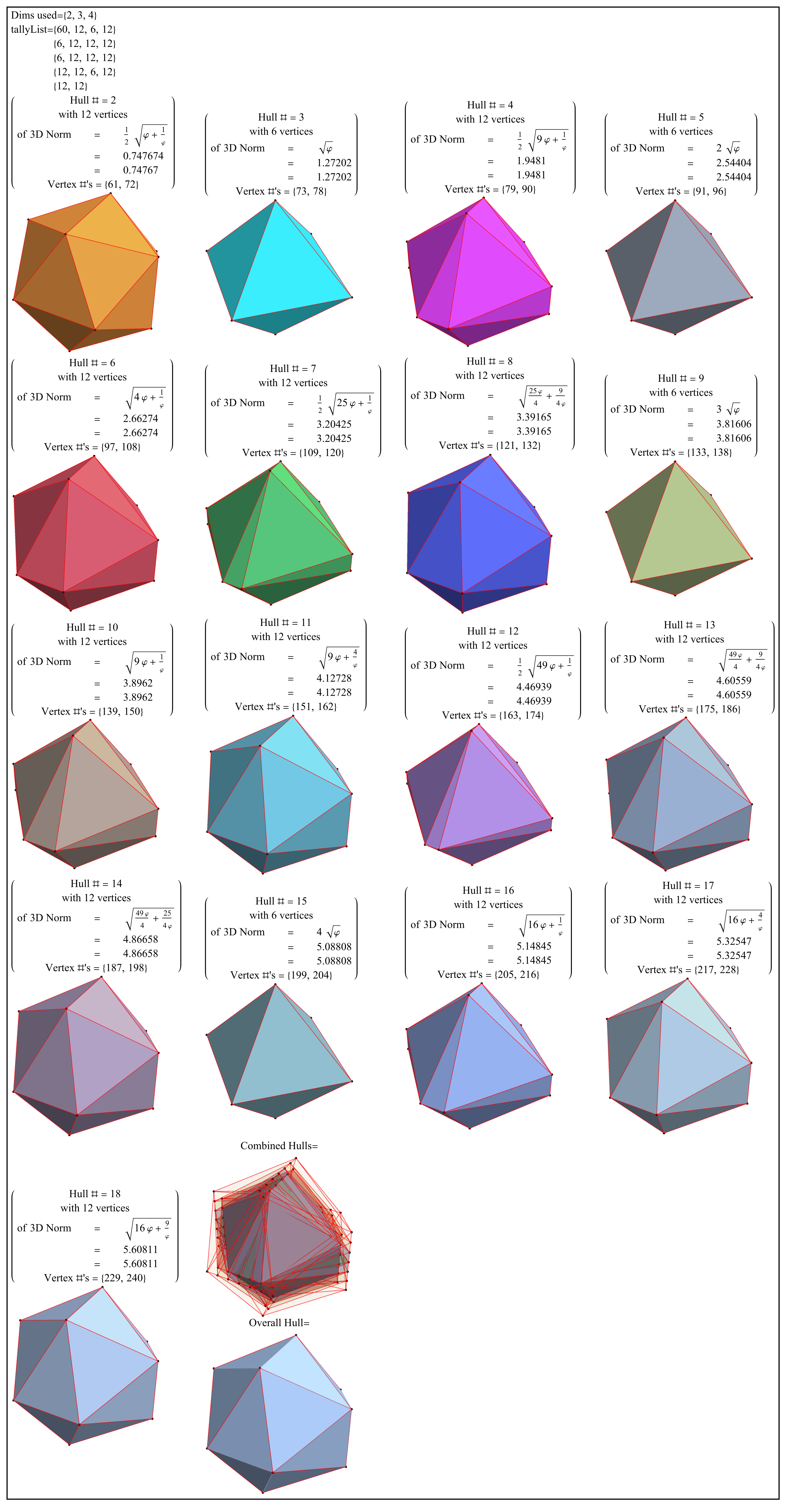}
\caption{\label{fig:UvertsInPerms3D-4}Orthogonal projection to 3D of the $\mathtt{cm\mathbb{U}}$-based vertex coordinates using dimensions \{2,3,4\}, noting the regular octahedral and irregular icosahedral hulls}
\end{figure}

\section{\label{app:B}\textit{Orthogonal projection to 3D of the $\mathtt{cm\mathbb{U}}$-based vertex coordinates using dimensions \{2,3,4\}}\\
Fig. \ref{fig:UvertsInPerms3D-4}
\ \\}

\end{document}